\documentclass[11pt]{article}

\usepackage{amsmath}

\begin{document}

\begin{center}
{\bf E.M. Ovsiyuk\footnote{e.ovsiyuk@mail.ru}\\[4mm]
  SPIN 1 FIELD IN THE LOBACHEVSKY SPACE $H_{3}$:\\HOROSPHERICAL COORDINATES, EXACT SOLUTIONS
  }

\vspace{2mm}
{\small  Mozyr State Pedagogical University named after I.P. Shamyakin, Belarus}

\end{center}

\date{}

\begin{abstract}

A complete system of solutions for a field with spin 1 in the space of constant negative curvature, Lobachevsky space $H_{3}$, has been constructed. The treatment is based on 10-dimensional Duffin--Kemmer formalism extended to curved model according to tetrad method by Tetrode--Weyl--Fock--Ivanenko, and specified in horospherical coordinates. The solving procedure substantially uses a generalized helicity operator. The Lobachevsky geometry acts along $z$ axis as a medium with simple reflecting properties. Restriction to massless case is performed as well.

\end{abstract}

The main task of the article is to construct a complete system of solutions
for  a field with spin 1 (massive and massless cases) in
 the space of constant negative curvature Lobachevsky space $H_{3}$; treatment is based on
 10-dimensional Duffin--Kemmer formalism extended to curved model according to
 tetrad method by Tetrode--Weyl--Fock--Ivanenko (see in  [1]). The main equation is
 $$
\left \{ \beta^{c} \left  [  i \hbar \; (\;  e_{(c)}^{\beta}
\partial_{\beta}  + {1\over 2} J^{ab} \gamma_{abc}  \; )\;
-\; {e \over c} A_{c}  \right ] \; - \;  mc  \right \} \Psi  = 0 \;.
\eqno(1)
$$

\noindent We will use horospherical coordinates and corresponding diagonal tetrad
$$
x^{a} = (t , r, \phi, z) \; , \qquad dS^{2}= dt^{2} - e^{-2z} (
dr^{2} + r^{2}d\phi^{2} ) - dz^{2} \; ,
$$
$$
e_{(a)}^{\beta} = \left | \begin{array}{cccc}
1 & 0 & 0 & 0 \\
0 & e^{z} & 0 & 0 \\
0 & 0 & {e^{z}\over r} & 0 \\
0 & 0 & 0 & 1
\end{array} \right | \; , \qquad
 e_{(a) \beta} = \left |
\begin{array}{llll}
1 & 0 & 0 & 0 \\
0 & -e^{-z} & 0 & 0 \\
0 & 0 & - re^{-z} & 0 \\
0 & 0 & 0 & - 1
\end{array} \right | \; .
\eqno(2a)
$$

\noindent For Christoffel symbols we have
 $ \Gamma^{0}_{\beta
\sigma} = 0 \; , \; \Gamma^{i}_{00} = 0 \; , \; \Gamma^{i}_{0j} =
0\;$,  and
$$
 x^{i}=(r,\phi,z)\; , \qquad
\Gamma^{i}_{\;\;jk } = g^{il} \; \Gamma_{l, jk}= {1 \over 2} \;
g^{i l} \; ( -\partial _{ l } \; g_{jk} +
\partial _{j } \; g_{lk } + \partial_{k} \; g_{lj } ) \; ,
$$
$$
\Gamma^{r}_{\;\; jk} = \left | \begin{array}{ccc}
0 & 0 & -1 \\
0 & -r & 0 \\
-1 & 0 & 0
\end{array} \right | \; ,
\qquad \Gamma^{\phi}_{\;\; jk} = \left | \begin{array}{ccc}
0 & {1 \over r }& 0 \\
{1\over r} & 0 & -1 \\
0 & -1 & 0
\end{array} \right | \; ,
$$
$$
\Gamma^{z}_{\;\; jk} = \left | \begin{array}{ccc}
e^{-2z} & 0 & 0 \\
0 & r^{2}e^{-2z} & 0 \\
0 & 0 & 0
\end{array} \right | \; .
\eqno(2b)
$$

\noindent Let us calculate covariant derivative for tetrad 4-vectors:
$$
A_{\beta ; \alpha }= {\partial A_{\beta} \over \partial
x^{\alpha}} - \Gamma^{\sigma} _{\alpha \beta} A_{\sigma} \;
\Longrightarrow
$$
$$
e_{(0) \beta ; \alpha }= {\partial e_{(0)\beta} \over \partial
x^{\alpha}} - \Gamma^{\sigma} _{\alpha \beta} e_{(0)\sigma} =
{\partial e_{(0)\beta} \over \partial x^{\alpha}} - \Gamma^{0}
_{\alpha \beta} \; e_{(0) 0} = 0 \; ,
$$
$$
e_{(1) \beta ; \alpha }= {\partial e_{(1)\beta} \over \partial
x^{\alpha}} - \Gamma^{r} _{\alpha \beta} \; e_{(1)r} = {\partial
e_{(1)\beta} \over \partial x^{\alpha}} + \Gamma^{r} _{\alpha
\beta} \; e^{-z} =
$$
$$
= \left | \begin{array}{cccc}
0 & 0 & 0 & 0 \\
0 & 0 & 0 & e^{-z} \\
0 & 0 & 0 & 0 \\
0 & 0 & 0 & 0
\end{array} \right | +
\left | \begin{array}{cccc}
0 & 0 & 0 & 0 \\
0 & 0 & 0 &- e^{-z} \\
0 & 0 & -r e^{-z} & 0 \\
0 & -e^{-z} & 0 & 0
\end{array} \right | =
 \left | \begin{array}{cccc}
0 & 0 & 0 & 0 \\
0 & 0 & 0 & 0 \\
0 & 0 & -r e^{-z} & 0 \\
0 & -e^{-z} & 0 & 0
\end{array} \right | ,
$$
$$
e_{(2) \beta ; \alpha }= {\partial e_{(2)\beta} \over \partial
x^{\alpha}} - \Gamma^{\phi} _{\alpha \beta} \; e_{(2)\phi} =
{\partial e_{(2)\beta} \over \partial x^{\alpha}} + \Gamma^{\phi}
_{\alpha \beta} r e^{-z} =
$$
$$
= \left | \begin{array}{cccc}
0 & 0 & 0 & 0 \\
0 & 0 & 0 & 0 \\
0 & - e^{-z}& 0 & r e^{-z} \\
0 & 0 & 0 & 0
\end{array} \right | +
\left | \begin{array}{cccc}
0 & 0 & 0 & 0 \\
0 & 0 & e^{-z} & 0 \\
0 & e^{-z} & 0 & -r e^{-z} \\
0 & 0 & -r e^{-z} & 0
\end{array} \right | = \left | \begin{array}{cccc}
0 & 0 & 0 & 0 \\
0 & 0 & e^{-z} & 0 \\
0 & 0 & 0 & 0 \\
0 & 0 & -r e^{-z} & 0
\end{array} \right | ,
$$
$$
e_{(3) \beta ; \alpha }= {\partial e_{(3)\beta} \over \partial
x^{\alpha}} - \Gamma^{z} _{\alpha \beta} \; e_{(3) z } = 0 +
\Gamma^{z} _{\alpha \beta} \;\; \Longrightarrow \;\; e_{(3) \beta
; \alpha } = \left | \begin{array}{cccc}
0 & 0 & 0 & 0 \\
0 & e^{-2z} & 0& 0 \\
0 & 0 &r^{2}e^{-2z} & 0 \\
0 & 0 & 0 &0
\end{array} \right | \; .
$$

\noindent Now we are ready to find Ricci rotation coefficients
$
\gamma_{ab c} = e_{(a)}^{\;\; \; \beta} \; e_{(b)\beta; \alpha} \;
e_{(c)}^{\alpha}$,  so we get
$$
\gamma_{ab 0} = e_{(a)}^{\;\;\; \beta} \; e_{(b)\beta; t } \;
e_{(0)}^{t} = 0 \; , \qquad
\gamma_{ab 1} = e_{(a)}^{\;\; \; a } \; e_{(b)a; r } \;
e_{(1)}^{r} \; ,
$$
$$
\gamma_{ab 2} = e_{(a)}^{\;\; \; \beta} \; e_{(b)\beta; \phi} \;
e_{(2)}^{\phi}\; , \qquad
\gamma_{ab 3} = e_{(a)}^{\;\; \; \beta} \; e_{(b)\beta; z } \;
e_{(3)}^{z} \; .
$$

\noindent Thus we obtain
\begin{eqnarray}
\gamma_{01 1}= \gamma_{02 1}= \gamma_{03 1} = 0 \; , \qquad
\gamma_{012}= \gamma_{022}= \gamma_{032} = 0 \; , \qquad
\gamma_{01 3}= \gamma_{02 3}= \gamma_{03 3} = 0 \; , \nonumber
\end{eqnarray}

\noindent and
$$
\gamma_{23 1} = e^{\phi}_{(2)} \; e_{(3)\phi; r} \; e^{r} _{(1)} =
0 \; ,
$$
$$
 \gamma_{31 1} = e^{z}_{(3)} \; e_{(1)z; r} \; e^{r}
_{(1)} = -1\; ,
$$
$$
 \gamma_{12 1} = e^{r}_{(1)} \; e_{(2)r; r}
\; e^{r} _{(1)} = 0 \; ,
$$
$$
\gamma_{23 2} = e_{(2)}^{\;\; \; \phi} \; e_{(3) \phi;\phi} \;
e_{(2)}^{\phi} = 1 \; ,
$$
$$
\gamma_{31 2} = e_{(3)}^{\;\; \; z} \; e_{(1) z ; \phi} \;
e_{(2)}^{\phi} = 0 \; ,
$$
$$
 \gamma_{12 2} = e_{(1)}^{\;\; \; r} \; e_{(2)r ; \phi} \; e_{(2)}^{\phi}
= {e^{z}\over r} \; ,
$$
$$
\gamma_{23 3} = e_{(2)}^{\;\; \; \phi} \; e_{(3)\phi ; z } \;
e_{(3)}^{z} = 0 \; ,
$$
$$
 \gamma_{31 3} = e_{(3)}^{\;\; \; z }
\; e_{(1) z; z } \; e_{(3)}^{z} = 0\; ,
$$
$$
 \gamma_{12 3} =
e_{(1)}^{\;\; \; r} \; e_{(2)r; z } \; e_{(3)}^{z} = 0 \; .
\eqno(3)
$$

Taking into account the above tetrad and Ricci coefficients we find 
an explicit representation for Duffin--Kemmer equation
(1)
$$
\left [   i \beta^{0}\partial_{t} +   e^{z}  [  i \beta^{1}
\;\partial_{r} + { i \beta^{2} \over r } (
\partial_{\phi} + J^{12} )  ] +
 i \beta^{3} \partial_{z}
+ i  \beta ^{1}  J^{13}  + i \beta^{2}  J^{23}-M
     \right ] \Psi = 0\,.
\eqno(4)
$$

\noindent
To separate the variable we use the following substitution
$$
\Psi = e^{-i\epsilon t  }  e^{im\phi}   \left |
\begin{array}{c}
\Phi_{0}  (r,z) \\
\vec{\Phi}(r,z)  \\
\vec{E} (r,z)  \\
\vec{H} (r,z)
\end{array} \right |;
\eqno(5)
$$

\noindent then consider eq. (4) in a block form
$$
 \left  [ \;  \epsilon \;   \left | \begin{array}{rrrr}
 0       &   0        &  0  &  0 \\
 0  &  0       &  i  & 0  \\
  0  &   -i       &   0  & 0\\
   0  &  0       &   0  & 0
\end{array} \right |    + i\; e^{z} \left |
\begin{array}{rrrrr}
  0       &  0       &    e_{1}  & 0       \\
    0   &  0       &   0      & \tau_{1} \\
   -e_{1}^{+}  &  0       &   0      & 0       \\
   0       &  -\tau_{1}&   0      & 0
\end {array} \right | {\partial \over \partial r }  \right.-
$$
$$
 - {e^{z} \over  r  } \; \left |
\begin{array}{rrrrr}
  0       &  0       &    e_{2}  & 0       \\
    0   &  0       &   0      & \tau_{2} \\
   -e_{2}^{+}  &  0       &   0      & 0       \\
   0       &  -\tau_{2}&   0      & 0
\end {array} \right |
 (   m     -  S_{3})
   + i \;   \left |
\begin{array}{rrrrr}
  0       &  0       &    e_{3}  & 0       \\
    0   &  0       &   0      & \tau_{3} \\
   -e_{3}^{+}  &  0       &   0      & 0       \\
   0       &  -\tau_{3}&   0      & 0
\end {array} \right | {\partial \over \partial z}+
$$
$$
\left. + i \; \left | \begin{array}{cccc}
0 & 0 &  -2 e_{3}  & 0 \\
0 & 0 & 0 &  - \tau_{3}  \\
0 & 0 &  0   & 0 \\
0 & + \tau_{3} & 0 & 0
\end{array} \right |
  - M\;   \right  ]  \left | \begin{array}{c}
\Phi_{0} \\
\vec{\Phi} \\
\vec{E} \\
\vec{H}
\end{array} \right |
 = 0 \;  ,
$$
$$
\eqno(6a)
$$

\noindent or
$$
   i \,e^{z}\,e_{1}  \partial _{r } \vec{E}
   -  {e^{z} \over   r} \; e_{2} (m  -S_{3} )  \vec{E}
    + i  (  \partial _{z }  - 2  ) e_{3}\vec{E}= M   \; \Phi_{0} \; ,
$$
$$
i \epsilon \; \vec{E} + i\,e^{z}\, \tau_{1} \partial_{r} \vec{H} -
e^{z}\,{\tau_{2} \over  r } ( \;  m  - S_{3} \;) \vec{H}  + i  (
\partial _{z }   - 1 ) \tau_{3}\vec{H} =M \; \vec{\Phi} \; ,
$$
$$
-i\epsilon \;  \vec{\Phi}  -i\,e^{z}\, e_{1}^{+} \partial_{r}
\Phi_{0} + e^{z}\,{m \over   r} \,e_{2}^{+} \Phi_{0}  - i \;
  e_{3}^{+}\partial _{z }
\Phi_{0} = M \;  \vec{E}\; ,
$$
$$
-i\,e^{z}\, \tau_{1} \partial_{r} \vec{\Phi} + e^{z}\,{ (m -
S_{3}) \over  r} \,\tau_{2}  \vec{\Phi}  - i (\partial _{z } - i )
\tau_{3}\vec{\Phi}= M \; \vec{H}\; . \eqno(6b)
$$

\noindent Performing calculation needed we get the system of 10 equations
$$
  \gamma  \,e^{z}\,(  { \partial  E_{1} \over \partial  r} - {\partial E_{3} \over \partial r
})  -  e^{z}\,{\gamma \over   r }   \left [  (m - 1 ) E_{1} + (m
+1) E_{3}   \right ]   -
$$
$$
-({\partial  \over \partial z} - 2  ) E_{2}
 = M \;   \Phi_{0} \; ,
\eqno(7a)
$$
$$
+i \epsilon \;    E_{1}    +   i \gamma \,e^{z}\, {\partial H_{2}
\over \partial  r}
 +    i e^{z}\,\gamma  {m  \over  r }   H_{2}
 + i  (   {\partial   \over \partial  z}- 1 ) H_{1}= M \;  \Phi_{1}\;,
$$
$$
+i \epsilon  \;   E_{2}  +  i \gamma \,e^{z}\, ( {\partial  H_{1}
\over \partial  r } + {\partial  H_{3} \over \partial  r } ) -
e^{z}\, {i\gamma \over   r } \left [  (m - 1)  H_{1}  -   (m + 1)
H_{3}  \right ]  = M \;  \Phi_{2}\;,
$$
$$
+i \epsilon  \;   E_{3}    +  i \gamma\,e^{z}\, {\partial   H_{2}
\over \partial r}  -   i e^{z}\,\gamma  {m  \over  r }  H_{2} - i
( {\partial  \over \partial z} - 1) H_{3} = M \;  \Phi_{3}
\eqno(7b)
$$
$$
-i  \epsilon  \;  \Phi_{1} +  \gamma \,e^{z}\,  {\partial \Phi_{0}
\over d r}   + e^{z}\, \gamma  {m \over    r }   \Phi_{0} = M \;
E_{1}\;,
$$
$$
-i \epsilon   \Phi_{2}  -   {\partial  \Phi_{0} \over \partial  z}
= M E_{2}\;,
$$
$$
-i \epsilon \;  \Phi_{3}  -   \gamma  \,e^{z}\, {\partial \Phi_{0}
\over \partial r}  + e^{z}\, \gamma {m   \over   r } \Phi_{0} = M
\;   E_{3}\;, \eqno(7c)
$$
$$
-i  \gamma  \,e^{z}\, {\partial  \Phi_{2} \over \partial  r}  -  i
e^{z}\,\gamma   {m   \over   r }  \Phi_{2}- i ( {\partial  \over
\partial  z} -1 ) \Phi_{1} = M \;   H_{1}\;,
$$
$$
 - i  \gamma \,e^{z}\, (  {\partial  \Phi_{1} \over \partial r } + { \partial \Phi_{3} \over \partial r })
+  {i e^{z}\,\gamma \over   r }  [ (m - 1) \Phi_{1} - (m+
1)\Phi_{3}  ]  = M  \;  H_{2} \; ,
$$
$$
-i  \gamma  \,e^{z}\, {\partial  \Phi_{2} \over \partial r}  +
 i\,e^{z}\, \gamma {m \over    r }   \Phi_{2}
  +i ({ \partial   \over \partial  z}- 1) \Phi_{3}=  M
\;  H_{3}\;. \eqno(7d)
$$

With the use of notation
$$
\gamma \;  ( {\partial  \over \partial r} +  { m - 1  \over   r }
) = a_{-} , \qquad \gamma \;  ( {\partial  \over \partial r} +  {
m + 1  \over   r } ) = a_{+}\;, \qquad \gamma \;  ( {\partial
\over \partial r} +  { m   \over   r } ) = a \; ,
$$
$$
\gamma \;  (- {\partial  \over \partial r} +  { m - 1  \over  r }
) = b_{-} , \qquad \gamma \;  (- {\partial  \over \partial r} +  {
m + 1  \over   r } ) = b_{+}\;, \qquad \gamma \;  (- {\partial
\over \partial r} +  { m   \over  r } ) = b \; ,
$$
$$
\eqno(8)
$$

\noindent they are written shorter
$$
  -e^{z}\,b_{-}  \; E_{1}   -e^{z}\, a_{+}  \;  E_{3}
-    ( {\partial  \over
\partial z } -2 ) E_{2}
 = M   \; \Phi_{0} \; ,
\eqno(9a)
$$
    $$
        i \,e^{z}\,  a  \;    H_{2} +i \epsilon   \,  E_{1}
 + i  \, ({\partial     \over \partial z}- 1) H_{1}  = M  \;    \Phi _{1}\;,
$$
$$
   - i \,e^{z}\, b_{-}  \;  H_{1}    +  i \,e^{z}\, a_{+} \; H_{3}
    + i \epsilon \;     E_{2}   = M  \;    \Phi_{2}\;,
    $$
$$
  - i \,e^{z}\, b\;  H_{2} +i \epsilon  \;    E_{3}
- i ( {\partial     \over \partial z} -  1) H_{3}  = M    \;
\Phi_{3}\; , \eqno(9b)
$$
$$
   e^{z}\, a \; \Phi_{0}  - i  \epsilon  \;    \Phi_{1} = M \;     E_{1}\;,
$$
$$
-i \epsilon    \Phi_{2}   -   {\partial  \over \partial z}  \Phi
_{0} = M  \;   E_{2} \;,
$$
$$
 e^{z}\,   b \;
\Phi_{0}  -i \epsilon  \;\Phi _{3}  = M    \;  E_{3}\;, \eqno(9c)
$$
$$
-  i  \,e^{z}\,a\;  \Phi_{2} \;- i \; ({\partial
 \over \partial z}  - 1) \Phi_{1} = M  \;   H_{1}\;,
$$
$$
  i  \,e^{z} \; b_{-}  \; \Phi_{1}
 - i\,e^{z}\,  a_{+} \; \Phi_{3}     = M  \;   H_{2} \; ,
$$
$$
  i \,e^{z} \; b  \;  \Phi_{2}
  + i \;  ({\partial    \over \partial z} - 1 )  \Phi_{3} =  M  \;    H_{3}\;.
\eqno(9d)
$$

To solve the problem let us consider an additional operator,  extended helicity operator
in the Lobachevsky space  (in  \cite{2,3,4} see more on the use of such  operators in
  the space $H_{3}$)
$$
 \Sigma \Psi = \sigma \Psi,\qquad
 \Psi = e^{-i\epsilon t  }  e^{im\phi}   \left |
\begin{array}{c}
\Phi_{0}  (r,z) \\
\vec{\Phi}(r,z)  \\
\vec{E} (r,z)  \\
\vec{H} (r,z)
\end{array} \right | ,
$$
$$
\left [\, e^{z}\, \left ( S_{1} \; {\partial  \over \partial r } +
iS_{2}
 { m - S_{3}   \over  r }\right )
+  ( {\partial \over \partial z} - 1)
 \;S_{3}
  \right ] \left |
\begin{array}{c}
\Phi_{0}   \\
\vec{\Phi}  \\
\vec{E}   \\
\vec{H}
\end{array} \right |= \sigma \;   \left |
\begin{array}{c}
\Phi_{0}   \\
\vec{\Phi}  \\
\vec{E}   \\
\vec{H}
\end{array} \right |\,.
\eqno(10)
 $$

\noindent From (10) it follows a system of 10 equations
 ($\gamma =1 /
\sqrt{2}$):
$$
0= \sigma \; \Phi_{0}\,, \eqno(11a)
$$
$$
  \gamma \,e^{z}\,  {\partial  \over \partial  r }   \Phi_{2}  +\gamma  \,e^{z}\,  { m   \over    r }  \Phi_{2}
+  ({ \partial  \over \partial  z} -1 )\,  \Phi_{1} = \sigma\,
\Phi_{1} \;,
$$
$$
  \gamma \,e^{z}\,  (  {\partial  \over \partial  r }   \Phi_{1}  + {\partial  \over \partial  r } \Phi_{3} )
-  \,e^{z}\, {\gamma \over   r } \, [ (m -  1) \Phi_{1} - (m + 1
)\Phi_{3}  ]  = \sigma  \,  \Phi_{2} \; ,
$$
$$
  \gamma   \,e^{z}\,{\partial  \over \partial  r }   \Phi_{2}  -
  \gamma\,e^{z}\, {m  \over  r }   \Phi_{2}
  -  ({ \partial  \over \partial  z} - 1)   \Phi_{3}=  \sigma \;
 \Phi_{3}\;,
\eqno(11b)
$$

$$
  \gamma \,e^{z}\,   {\partial  \over \partial  r }   E_{2}  +\gamma \,e^{z}\,   { m   \over    r }  E_{2}
+ ({ \partial  \over \partial  z} - 1 )  E_{1} = \sigma\, E_{1}
\;,
$$
$$
  \gamma \,e^{z}\,  (  {\partial  \over \partial  r }   E_{1}  + {\partial  \over \partial  r } E_{3} )
- \,e^{z}\,  {\gamma \over  r }  \,[ (m - 1) E_{1} - (m + 1 )E_{3}
]  = \sigma  \,   E_{2} \; ,
$$
$$
  \gamma  \,e^{z}\,  {\partial  \over \partial  r }   E_{2}  -
  \gamma \,e^{z}\,  {m  \over    r }   E_{2}
  -  ({ \partial  \over \partial  z} - 1 )  E_{3}=  \sigma \;   E_{3}\;,
\eqno(11c)
$$

$$
  \gamma  \,e^{z}\,  {\partial  \over \partial  r }   H_{2}  +\gamma \,e^{z}\,   { m   \over    r }  H_{2}
+ ({ \partial  \over \partial  z} - 1 )  H_{1} = \sigma\,  H_{1}
\;,
$$
$$
  \gamma \,e^{z}\,  (  {\partial  \over \partial  r }   H_{1}  + {\partial  \over \partial   r } H_{3} )
- \,e^{z}\,  {\gamma \over   r }\,  [ (m - 1) H_{1} - (m + 1
)H_{3}  ]  = \sigma  \,  H_{2} \; ,
$$
$$
  \gamma \,e^{z}\,   {\partial  \over \partial  r }   H_{2}  -
  \gamma \,e^{z}\,  {m  \over   r }   H_{2}
  -  ({ \partial  \over \partial  z} - 1 )  H_{3}=  \sigma    \;H_{3}\;.
\eqno(11d)
$$

\noindent It can be presented shorter
$$
0= \sigma \; \Phi_{0}\,,
\eqno(12a)$$
$$
 ({ \partial  \over \partial  z} -1 )  \Phi_{1} =
\sigma\,  \Phi_{1} - a \,e^{z}\, \Phi_{2}  \;,
$$
$$
- \,e^{z}\, b_{-} \Phi_{1} + \,e^{z}\, a_{+} \Phi_{3} = \sigma \;
\Phi_{2}
  \; ,
$$
$$
   -  ({ \partial  \over \partial  z} - 1 )   \Phi_{3}=
\sigma \;    \Phi_{3} +  b\; \,e^{z}\,   \Phi_{2} \;, \eqno(12b)
$$
$$
 ({ \partial  \over \partial  z} - 1 )  E_{1} =
\sigma \;  E_{1} -a\,e^{z}\,     E_{2}  \;,
$$
$$
-\,e^{z}\,  b_{-} E_{1} + \,e^{z}\, a_{+} E_{3} = \sigma \; E_{2}
  \; ,
$$
$$
     - ({ \partial  \over \partial  z} -
1 )   E_{3}=  \sigma \;   E_{3} + b\;   \,e^{z}\,  E_{2}\;,
\eqno(12c)
$$
$$
 ({ \partial  \over \partial  z} - 1 )  H_{1} =
 \sigma\, H_{1} -  a\;  \,e^{z}\,   H_{2}\;,
$$
$$
- \,e^{z}\, b_{-} H_{1} +\,e^{z}\,  a_{+} H_{3} = \sigma\;   H_{2}
  \; ,
$$
$$
    - ({ \partial  \over \partial  z} -1)    H_{3}=
\sigma \;  H_{3} + b\;  \,e^{z}\,   H_{2} \;. \eqno(12d)
$$

Substituting (12) into (9), we arrive at much more simple equations
 (note that  $(13b)$  and  $(13d)$  provide us with linear restrictions)
$$
 -e^{z}\,b_{-}  \; E_{1}   -e^{z}\, a_{+}  \;  E_{3}
-    ( {\partial  \over
\partial z } -2 ) E_{2}
 = M   \; \Phi_{0} \; ,
\eqno(13a)
$$
    $$
        i \epsilon  \;  E_{1}
 + i  \sigma\,  H_{1}  = M  \;  \Phi _{1}\;,
$$
$$
       i \epsilon \;    E_{2}  + i \sigma\;  H_{2}   = M  \; \Phi_{2}\;,
    $$
$$
  i \epsilon  \;  E_{3}
+ i \sigma \;  H_{3}   = M    \; \Phi_{3}\; , \eqno(13b)
$$
$$
   e^{z}\, a \; \Phi_{0}  - i  \epsilon  \;    \Phi_{1} = M \;     E_{1}\;,
$$
$$
-i \epsilon    \Phi_{2}   -   {\partial  \over \partial z}  \Phi
_{0} = M  \;   E_{2} \;,
$$
$$
 e^{z}\,   b \;
\Phi_{0}  -i \epsilon  \;\Phi _{3}  = M    \;  E_{3}\;, \eqno(13c)
$$
$$
-   \sigma\,  \Phi_{1}   = M  \; H_{1}\;,
$$
$$
  - i \sigma \;  \Phi_{2}    = M  \;   H_{2} \; ,
$$
$$
 -i\sigma \;    \Phi_{3}   =  M  \;    H_{3}\;.
\eqno(13d)
$$

In the system  (12), one can note three groups of similar equations (in essence, they are much the same).
Let us detail  eqs.   $(12a)$ for $\Phi_{j}(z,r)$
$$
 ({ \partial  \over \partial  z} -1 )  \Phi_{1} =
\sigma\,  \Phi_{1} - a \,e^{z}\, \Phi_{2}  \;,
$$
$$
- \,e^{z}\, b_{-} \Phi_{1} + \,e^{z}\, a_{+} \Phi_{3} = \sigma \;
\Phi_{2}
  \; ,
$$
$$
   -  ({ \partial  \over \partial  z} - 1 )   \Phi_{3}=
\sigma \;    \Phi_{3} +  b\; \,e^{z}\,   \Phi_{2} \;. \eqno(14a)
$$

\noindent By substitutions
$$
\Phi_{1} = e^{z} \varphi_{1}(r,z) \;, \qquad  \Phi_{2} = e^{2z}
\varphi_{2}(r,z) \;, \qquad \Phi_{3} = e^{z} \varphi_{3}(r,z)
 \eqno(14b)
$$

\noindent the system  $(14a)$ is simplified to
$$
 a\;    \varphi_{2}  = e^{-2z} (+ \sigma -  { \partial  \over \partial  z}  ) \;  \varphi_{1} \;,
$$
$$
- b_{-} \varphi_{1} + a_{+} \varphi_{3} = \sigma\;   \varphi_{2}
  \; ,
$$
$$
 b\;    \varphi_{2}    = e^{-2z} ( -\sigma -  { \partial  \over \partial  z}) \;   \varphi_{3}\; .
\eqno(14c)
$$

\noindent With the use of differential operators
$b_{-}$ and $a_{+}$  (see (8)), let us introduce new variables
$$
b_{-} \varphi_{1} = \bar{\varphi}_{1} \;, \qquad a_{+}\varphi_{3}
= \bar{\varphi}_{3} \; ; \eqno(15a)
$$

\noindent then from   $(14c)$ it follows
$$
 b_{-} a\;    \varphi_{2}    = e^{-2z}( +\sigma -{ \partial  \over \partial  z})    \bar{\varphi}_{1}    \;,
$$
$$
\bar{\varphi}_{3} - \bar{\varphi}_{1}   = \sigma\;   \varphi_{2}
  \; ,
$$
$$
 a_{+} b\;    \varphi_{2}         = e^{-2z} (- \sigma -{ \partial  \over \partial  z})  \;  \bar{\varphi}_{3}  \; .
\eqno(15b)
$$

\noindent Note that the first and the third equations involve one the same operator
$$
b_{-}a =a_{+}b = {1 \over 2} \left ( -{\partial ^{2} \over
\partial r^{2}} - {1 \over  r} {\partial \over \partial
r} + {m^{2} \over r^{2} } \right ) = \Delta \; . \eqno(15c)
$$

First, let us assume $\sigma \neq 0$. From the first ant the third equations in $(15b)$ it follows
$$
 \sigma ( \bar{\varphi}_{1}  + \bar{\varphi}_{3} )   =
-{ \partial  \over \partial  z}   (\bar{\varphi}_{3}  -
\bar{\varphi}_{1} ) =- \sigma { \partial  \over \partial  z}
\varphi_{2} \; . \eqno(16a)
$$

\noindent Thus, we derive two simple relations
$$
  \bar{\varphi}_{3}  + \bar{\varphi}_{1}    =  - { \partial  \over \partial  z} \varphi_{2}\;, \qquad
\bar{\varphi}_{3}   - \bar{\varphi}_{1}    = \sigma\; \varphi_{2}
  \; ;
$$

\noindent which  provide us with expressions for
 $\bar{\varphi}_{1} $ and
$\bar{\varphi}_{3} $  through  $\varphi_{2}$
$$
\bar{\varphi}_{3}  = {1 \over 2}(+\sigma -{ \partial  \over
\partial  z}) \varphi_{2} \; , \qquad \bar{\varphi}_{1}  = {1
\over 2}(-\sigma -{ \partial  \over \partial  z}) \varphi_{2} \; .
\eqno(16b)
$$

\noindent In turn, substitution them into the first and the third equations in
$(15b)$  we  get one the same equation for $\varphi_{2}$
$$
 b_{-} a\;    \varphi_{2}    =e^{-2z} ( \sigma -{ \partial  \over \partial  z})
  {1 \over 2}(-\sigma -{ \partial  \over \partial  z}) \varphi_{2}    \;,
$$
$$
 a_{+} b\;    \varphi_{2}         = e^{-2z}
(- \sigma -{ \partial  \over \partial  z})  \; {1 \over 2}(+\sigma
-{ \partial  \over \partial  z}) \varphi_{2}
 \; .
\eqno(17a)
$$

In eq.   $(17a)$, one can separate variables
$$
\varphi_{2}(r,z)  =\varphi_{2}(r) \;  \varphi_{2}(z)\;,
$$
$$
{1 \over h_{2}(r)}  \; (2 \Delta  ) \;    h_{2}(r)    = {1 \over
h_{2}(z)} e^{-2z} ( {d^{2}   \over d  z^{2} } - \sigma ^{2} )
   \varphi_{2}  (z) = \Lambda  \; ;
$$

\noindent who we get two differential equations, in $r$ and $z$ variables
respectively:
$$
 2\Delta  \;    \varphi_{2} (r) = \Lambda \; \varphi_{2} (r) \; ,
\eqno(17b)
$$
$$
  ({d^{2}   \over d  z^{2} } - \sigma ^{2} )
   \varphi_{2}(z)   =  \Lambda  e^{+2z}  \;  \varphi_{2} (z) \; .
\eqno(17c)
$$

\noindent By  $\varphi_{2} (r,z)  =\varphi_{2}(r)  \varphi_{2}(z)$
one can calculate two remaining components
$$
\bar{\varphi}_{1}  = {1 \over 2}(-\sigma -{ \partial  \over
\partial  z}) \varphi_{2} \; , \qquad \bar{\varphi}_{3}  = {1
\over 2}(+\sigma -{ \partial  \over \partial  z}) \varphi_{2} \; .
\eqno(17d)
$$

Similar analysis gives
$$
 2 \Delta  \;    e_{2} (r) = \Lambda \; e_{2} (r) \; ,
$$
$$
  ({d^{2}   \over d  z^{2} } - \sigma ^{2} )
   e_{2}(z)   =  \Lambda e^{+2z} \;  e_{2} (z) \; ,
$$
$$
\bar{e}_{1}  = {1 \over 2}(-\sigma -{ \partial  \over \partial z})
e_{2} \; ,\qquad \bar{e}_{3}  = {1 \over 2}(+\sigma -{
\partial  \over \partial  z}) e_{2} \; ;
\eqno(18a)
$$

$$
 2 \Delta  \;    h_{2} (r) = \Lambda \; h_{2} (r) \; ,
$$
$$
  ({d^{2}   \over d  z^{2} } - \sigma ^{2} )
   h_{2}(z)   =  \Lambda e^{+2z} \;  h_{2} (z) \; ,
$$
$$
\bar{h}_{1}  = {1 \over 2}(-\sigma -{ \partial  \over \partial z})
h_{2} \; , \qquad \bar{h}_{3}  = {1 \over 2}(+\sigma -{
\partial  \over \partial  z}) h_{2} \; .
\eqno(18b)
$$

In explicit form eq. $(17b)$ is
$$
\left[{d^{2}\over dr^{2}}+{1\over r}\,{d\over dr}-{m^{2}\over r^{2}}+\Lambda\right]\varphi_{2}=0
$$

\noindent in  the new variable  $y=\sqrt{\Lambda}\;r$ it is the Bessel equation
$$
\left ({d^{2}\over d y^{2}}+{1\over y}\,{d\over d y}+1-{m^{2}\over y^{2}}\right )\varphi_{2}=0
$$

\noindent with solutions
$$
\varphi_{2} (x) =A\,J_{\,m}(y)+B\,Y_{m}(y)\,.
\eqno(19a)
$$

In turn, eq. $(17c)$ in the variable
 $x=i\,\sqrt{\Lambda}\,e^{+z}$ is the Bessel equation as well
 $$
\left ( {d^{2}\over d x^{2}}+{1\over x}\,{d\over d x}+1-{\sigma^{2}\over x^{2}}\right )\varphi_{2}=0\,,
$$

\noindent with solutions
$$
\varphi_{2}=C\,J_{\,\sigma}(x)+D\,Y_{\sigma}(x)\,.
\eqno(19b)
$$

Now let us turn to the system $(15b)$  for the  case
$\sigma =0$:
$$
\bar{\varphi}_{3} = \bar{\varphi}_{1} = \bar{\varphi}   \; ,
$$
$$
 \Delta   \;  \varphi_{2}    = - e^{-2z}  { \partial  \over \partial  z}    \bar{\varphi}    \;,
$$
$$
 \Delta \;    \varphi_{2}         = -  e^{-2z} { \partial  \over \partial  z}  \;  \bar{\varphi} \; .
\eqno(20)
$$

\noindent Similar equations take place for $E_{j},\, H_{j}$.
Now the system (13) becomes simpler
$$
 -e^{z}\,b_{-}  \; E_{1}   -e^{z}\, a_{+}  \;  E_{3}
-    ( {\partial  \over
\partial z } -2 ) E_{2}
 = M   \; \Phi_{0} \; ,
$$
    $$
        i \epsilon  \;  E_{j}
= M  \;  \Phi _{j}\;,
 $$
$$
   e^{z}\, a \; \Phi_{0}  - i  \epsilon  \;    \Phi_{1} = M \;     E_{1}\;,
$$
$$
-i \epsilon    \Phi_{2}   -   {\partial  \over \partial z}  \Phi
_{0} = M  \;   E_{2} \;,
$$
$$
 e^{z}\,   b \;
\Phi_{0}  -i \epsilon  \;\Phi _{3}  = M    \;  E_{3}\;,
$$
$$
 H_{j} = 0\;,
$$

\noindent or in other  variables
$$
 -2  \bar{e}  -     {\partial  \over
\partial z }  e_{2}
 = M   \;e^{-2z}  \Phi_{0} \; ,
$$
    $$
        i \epsilon  \;  \bar{e} = M  \;  \bar{\varphi} \;, \qquad
     i \epsilon  \;  e_{2}
= M  \;  \varphi _{2}\;,
$$
$$
-i \epsilon    e^{2z} \varphi_{2}   -   {\partial  \over \partial
z}  \Phi _{0} = M  \;   e^{2z} e_{2} \;,
$$
$$
    \Delta \;
\Phi_{0}  -i \epsilon  \;\bar{\varphi}   = M    \;  \bar{e} \;,
$$
$$
 H_{j} = 0\;.
\eqno(21)
$$

\noindent
Let us exclude  the variables $\bar{e}, e_{2}$:
    $$
 H_{j} =0, \qquad        \bar{e} = { M \over i \epsilon }  \;  \bar{\varphi} \;, \qquad
      \;  e_{2} = {M \over i \epsilon }  \;  \varphi _{2}\;,
$$
$$
 -2  { M \over i \epsilon }  \;  \bar{\varphi}   -     {\partial  \over
\partial z }   {M \over i \epsilon }  \;  \varphi _{2}
 = M   \;e^{-2z}  \Phi_{0} \; ,
$$
$$
-i \epsilon    e^{2z} \varphi_{2}   -   {\partial  \over \partial
z}  \Phi _{0} = M  \;   e^{2z} {M \over i \epsilon }  \;  \varphi
_{2} \;,
$$
$$
    \Delta \;
\Phi_{0}  -i \epsilon  \;\bar{\varphi}   = M    \;  { M \over i
\epsilon }  \;  \bar{\varphi} \;,
$$

\noindent or
$$
 -   2  \;  \bar{\varphi}   -     {\partial  \over
\partial z }     \varphi _{2}
 =  i \epsilon e^{-2z} \Phi_{0} \; ,
$$
$$
  \varphi _{2} =  {i \epsilon  \over (  \epsilon ^{2}  - M^{2} )} e^{-2z}    {\partial  \over \partial z}  \Phi_{0}\;,
$$
$$
   \bar{\varphi} =  { i\epsilon  \over (  M^{2} - \epsilon^{2} )  }   \Delta \; \Phi_{0}       \;.
\eqno(22)
$$

\noindent
From (22), allowing for
the second and the third equations, from the first we derive
$$
 -   2  \;   { i\epsilon  \over (  M^{2} - \epsilon^{2} )  }   \Delta \; \Phi_{0}
   -     {\partial  \over
\partial z }     {i \epsilon  \over (  \epsilon ^{2}  - M^{2} )} e^{-2z}    {\partial  \over \partial z}  \Phi_{0}
 =  i \epsilon  e^{-2z}  \Phi_{0} \; ,
$$

\noindent that is
$$
  \left [   2      \Delta
   -     {\partial  \over
\partial z }      e^{-2z}    {\partial  \over \partial z}
 -    (\epsilon^{2} - M^{2} ) e^{-2z}  \right ]  \Phi_{0} = 0  \; .
 \eqno(23a)
$$

Let it be
$\Phi_{0} (r,z) = \Phi_{0}(r)   \Phi_{0} (z)$, then the variables are separated
$$
   {1 \over  \Phi_{0}(r)}  2      \Delta \Phi_{0}(r)
   =     {1 \over \Phi_{0}(r)}  \left (  {\partial  \over
\partial z }      e^{-2z}    {\partial  \over \partial z}
 +    (\epsilon^{2} - M^{2} ) e^{-2z}  \right ) \Phi_{0}(z) = \Lambda  \,,
 \eqno(23b)
$$

\noindent and further we get
$$
{1 \over  \Phi_{0}(r)}  2      \Delta \Phi_{0}(r) = \Lambda \; ,
$$
$$
{1 \over \Phi_{0}(r)}  \left (  {\partial  \over
\partial z }      e^{-2z}    {\partial  \over \partial z}
 +    (\epsilon^{2} - M^{2} ) e^{-2z}  \right ) \Phi_{0}(z) = \Lambda \; .
 \eqno(23c)
 $$

Explicitly they read
$$
  \left ( {d ^{2} \over
d r^{2}} + {1 \over  r} {d \over d
r} - {m^{2} \over  r^{2}}  + \Lambda \,\right )\Phi_{0}(r)=0 \; ,
$$
$$
  \left (       {d^{2} \over d z^{2}}-2\, {d \over d z}
 +    (\epsilon^{2} - M^{2} )    -   \Lambda \,e^{+2z}\,\right ) \Phi_{0}(z)=0 \, .
 \eqno(23d)
 $$

\noindent With substitution
 $\Phi_{0}(z)=e^{z}\,\phi_{0}(z)$ we get from the second equation
 $$
 \left ({d^{2}\over dz^{2}} +(\epsilon^{2}-M^{2})  -1  -\Lambda\,e^{+2z}  \right )\phi_{0}(r)=0 \; ;
\eqno(24a)
$$

\noindent  in variable  $x=i\,\sqrt{\Lambda}\,e^{z}$ it reduce to the Bessel form
$$
\left({d^{2}\over d x^{2}}+{1\over x}\,{d \over d x}+1-{1-\epsilon^{2}+M^{2}\over x^{2}}\right)\,\phi_{0}=0
\eqno(24b)
$$

\noindent with solutions
$$
\phi_{0}=C_{1}\,J_{1-\epsilon^{2}-M^{2}}(x)+C_{2}\,Y_{1-\epsilon^{2}-M^{2}}(x)\,.
\eqno(24c)
$$

Concluding, let us specify the massless case.
Instead of (13) now we have
$$
 -e^{z}\,b_{-}  \; E_{1}   -e^{z}\, a_{+}  \;  E_{3}
-    ( {\partial  \over
\partial z } -2 ) E_{2}
 = 0 \; ,
$$
    $$
        i \epsilon  \;  E_{1}
 + i  \sigma\,  H_{1}  = 0\;,\;\;
  i \sigma\;  H_{2}
    + i \epsilon \;    E_{2}   = 0\;,\;\;
  i \epsilon  \;  E_{3}
+ i \sigma \;  H_{3}   = 0 \; ,
$$
$$
   e^{z}\, a \; \Phi_{0}  - i  \epsilon  \;    \Phi_{1} =     E_{1}\;,
$$
$$
-i \epsilon    \Phi_{2}   -   {\partial  \over \partial z}  \Phi
_{0} =    E_{2} \;,
$$
$$
 e^{z}\,   b \;
\Phi_{0}  -i \epsilon  \;\Phi _{3}  =   E_{3}\;,
$$
$$
-   \sigma\,  \Phi_{1}   = H_{1}\;,\qquad
  - i \sigma \;  \Phi_{2}    =    H_{2} \; , \qquad
 -i\sigma \;    \Phi_{3}   =      H_{3}\;.
\eqno(25)
$$

The most interesting is the case of  $\sigma=0$, the system (25) gives
    $$
         E_{j}  = 0\;, \qquad H_{j}=0 \; ,
$$
$$
    a \; \Phi_{0}  - i  \epsilon  \;    \varphi_{1} =     0\;,
$$
$$
-i \epsilon    \Phi_{2}   -   {\partial  \over \partial z}  \Phi
_{0} =    0 \;,
$$
$$
    b \;
\Phi_{0}  -i \epsilon  \;\varphi _{3}  =   0 \; ,
\eqno(26)
$$

\noindent
from whence it follows
    $$
         E_{j}  = 0\;, \qquad H_{j} = 0 \; ,
 \eqno(27a)
$$
$$
  \bar{\varphi}  = {1 \over i  \epsilon }   \Delta  \; \Phi_{0}      \;, \qquad
 \varphi_{2}   =  - {1 \over  i  \epsilon }  e^{-2z} {\partial  \over \partial z}  \Phi
_{0} =    0 \;,
\eqno(27b)
$$

\noindent
where $\Phi_{0}$ is an arbitrary function. This class describes solutions of gradient type with vanishing electromagnetic tensor.
\vspace{3mm}

Author is grateful   to V.M. Red'kov for encouragement  and advices.

\end{document}